\documentstyle[epsf,floats,prb,aps]{revtex}

\begin{document}
\draft

\twocolumn[\hsize\textwidth\columnwidth\hsize\csname
@twocolumnfalse\endcsname

\preprint{February 18, 1997}


\title{Ab-initio study of BaTiO$_3$ surfaces}

\author{J.~Padilla and David Vanderbilt}

\address{Department of Physics and Astronomy, 
Rutgers University, Piscataway, NJ 08855-0849}

\date{February 18, 1997}

\maketitle

\begin{abstract}
We have carried out first-principles total-energy calculations
of (001) surfaces of the tetragonal and cubic phases of BaTiO$_3$.
Both BaO-terminated (type I) and TiO$_2$-terminated (type II)
surfaces are considered, and the atomic configurations have been
fully relaxed.  We found no deep-gap surface states for any of
the surfaces, in agreement with previous theoretical studies.
However, the gap is reduced for the type-II surface, especially
in the cubic phase.  The surface relaxation energies are found
to be substantial, i.e., many times larger than the bulk
ferroelectric well depth.  Nevertheless, the influence of the
surface upon the ferroelectric order parameter is modest;
we find only a small enhancement of the ferroelectricity
near the surface.
\end{abstract}

\vskip2pc]

\narrowtext

\section{Introduction}

Recently there has been a surge of interest in the application
of first-principles density-functional calculations to the study
of the rich phenomenology of the perovskite oxides, with special
attention to ferroelectric (FE) properties.\cite{line} From these
investigations, it has been found that the FE instability in
these materials occurs as a result of a delicate balance between
long-range Coulomb interactions that favor the FE state, and
short-range forces that favor the cubic perovskite
phase.\cite{resta,zhong1} Moreover, the ferroelectric properties
are well know to degrade in thin-film\cite{tsai}
and particulate\cite{niepce} geometries,
suggesting that the FE state could be very sensitive to surface effects.

The cubic perovskites have the chemical formula ABO$_3$.  For
II-IV perovskites (e.g., BaTiO$_3$) A is a divalent cation and
B is a tetravalent transition metal, while for I-V perovskites
(e.g., KNbO$_3$) they are mono- and pentavalent respectively.
The (001) and (111) surfaces of these materials have been most
investigated experimentally.\cite{cox}  There are two possible
terminations of the (001) surface: the AO-terminated surface
(type-I) and the BO$_2$-terminated surface (type-II).  In II-IV
perovskites, the AO and BO$_2$ layers are charge-neutral, so
that both type-I and type-II surfaces are non-polar.  For I-V
perovskites, the corresponding surfaces are instead polar.  As
for (111) surfaces, the atomic planes in this direction are of
the form AO$_3$ and B, and are charged in either case,
so that the (111) surfaces are polar.  Since polar surfaces
are expected to be relatively unstable, we have chosen to focus
here on the (001) surfaces of a II-IV perovskite, BaTiO$_3$.

Due in part to the catalytic properties of SrTiO$_3$ and
BaTiO$_3$,\cite{tom} there has been a continuous interest in the
surface properties of these materials.  There have been previous
theoretical studies especially for the case of the paraelectric
SrTiO$_3$ surface, but also for BaTiO$_3$.  Wolfram and coworkers
\cite{wolf1}, using a linear combination of atomic orbitals (LCAO)
cluster method, predicted mid-gap surface states for SrTiO$_3$,
in disagreement with experimental investigations.
\cite{powell,henrich}  Only after {\it ad hoc} modifications to
this model could the experimental results be accounted
for.\cite{wolf2} Tsukada {\it et al.}\cite{suka} employed the DV
X$\alpha$ cluster method to study SrTiO$_3$ surfaces, finding no
mid-gap surface states.  However, cluster methods are not very
suitable for high-accuracy calculations of relaxations and
electronic states on infinite surfaces, underlining the need for
the application of more accurate, self-consistent techniques.
While such techniques have recently yielded a great deal of insight
into bulk perovskites,\cite{zhong1,wz2,cohen,singh} 
their application to the study of perovskite surfaces has not
been very extensive.  In fact, we only know of two such studies.
Cohen\cite{cohen1,cohen2} presented linearized augmented plane wave
(LAPW) calculations performed for slabs of tetragonal BaTiO$_3$
with (001) and (111) surfaces, using both symmetrical and
asymmetrical terminations.  Although some relaxations were allowed,
the atomic positions were not fully relaxed.  Kimura {\it et
al.}\cite{kimura} used a plane-wave ultrasoft-pseudopotential
\cite{vand2} approach (as in the present work) to study the
TiO$_2$-terminated (001) surface of SrTiO$_3$, with and without
oxygen vacancies at the surface.  Again, the slabs were not fully
relaxed.

In contrast, we study here symmetrically-terminated type-I and
type-II surfaces of tetragonal and cubic BaTiO$_3$ (001) for
which the coordinates have been fully relaxed by minimizing the
total energy.  This allows us to study the influence of surface
relaxation effects upon the FE distortion.  For the tetragonal
phase, we consider only the case of the tetragonal $c$ axis (i.e.,
polarization) parallel to the surface.  (Polarization normal to
the surface is suppressed by depolarization fields,\cite{zhong1}
at least for clean surfaces.)  We employed the
ultrasoft-pseudopotential method\cite{vand2} within the local-density
approximation (LDA).  This technique permits us to calculate the
Hellmann-Feynman forces on each atom, making it possible to find
the relaxed structure much more efficiently than for methods that
compute only total energies.

Experimental studies of perovskites surfaces are complicated by
the presence of surface defects,\cite{cord} making it difficult
to verify the surface stoichiometry.  Therefore, most experimental
investigations have not been very conclusive.  On SrTiO$_3$
surfaces, the situation is better: the surface relaxations have
been studied,\cite{hikita} and (as mentioned above) the absence
of mid-gap surface state has been demonstrated.\cite{powell} For
BaTiO$_3$ surfaces, the experimental reports seem less conclusive.
For example, evidence both for\cite{robey} and against\cite{chen}
surface gap states in this material have been published.

Regarding the degradation of FE properties for thin films and
small particles as mentioned above,\cite{tsai,niepce}
there does not seem to be any consensus about the origin of these
effects.  One possibility is that it is completely intrinsic, i.e.,
that the very presence of the surface suppresses the FE order in
the vicinity of the surface.  However, there are many other
possible causes.  These include the effects of surface-induced strain;
perturbations of the chemical composition near the surface
related to the presence of impurities, oxygen vacancies, or other
defects; and the depolarization fields for the case of particles
(or for films with polarization perpendicular to the surface).
Here, we take a modest step in the direction of sorting out these
effects by characterizing the purely intrinsic coupling between the
presence of a surface and the FE order, for the case of a free
(vacuum-terminated) surface.  As we shall see, we find the surface
relaxation energies are large compared to FE distortion energies.
However, we find very little effect for type-I surfaces, and
only a modest enhancement of the FE order at type-II surfaces,
with indications that it will be mainly confined to just the
first few atomic layers near the surface.  Thus, it appears
unlikely that intrinsic surface effects are responsible for the
degradation of FE order in thin films and particles.

The remainder of the paper is divided as follows.  In Sec. II, we
describe the technical aspects of our first-principles calculations.  
In Sec. III, we report the results of our simulations. Finally, the main
conclusions of the paper are reviewed in Sec. IV. 

\section{Theoretical Details}
\label{sec:theory}

We carried out self-consistent total-energy pseudopotential
calculations in which the  electronic wave functions were expanded
in a plane-wave basis.  The core electrons were frozen, and for a
given geometry of the ions, the valence electron wavefunctions were
obtained by minimizing the Kohn-Sham total-energy functional using
a conjugate-gradient technique.\cite{king}  The
exchange-correlation potential was treated with the LDA
approximation in the Ceperley-Alder form.\cite{ceper}  The forces
on each ion were relaxed to less than $0.02$ eV/\AA\ using a
modified Broyden scheme.\cite{vand1}

The Vanderbilt ultrasoft pseudopotential scheme\cite{vand2}
was employed.  In this approach, the the norm-conservation 
constraint is relaxed, allowing one to treat rather localized
orbitals with a modest plane-wave cutoff.  The pseudopotentials for
Ti, Ba and O are identical to those used previously\cite{king} in a
study of bulk perovskites.  The semicore Ti $3s$ and $3p$ states
and Ba $5s$ and $5p$ states are included as valence levels.  A
plane-wave cutoff of 25 Ry has been used throughout; previous work
has shown that the results are well-converged at this cutoff.

\begin{table}
\caption{Computed and experimental values of structural parameters
for BaTiO$_3$ in bulk cubic and tetragonal phases.  $a$ and $c$
are lattice constants; $\delta_{\rm x}$ are displacements associated
with the FE instability as a fraction of $c$.
O$_{\rm I}$ is the oxygen lying along $\hat{x}$ from the Ti atom,
and $\delta_{\rm x}$(O$_{\rm II})=\delta_{\rm x}$(O$_{\rm III})$
by symmetry).}
\vspace{0.1in}
\begin{tabular}{llrr}
Phase & Parameter &Exper.\tablenotemark[1] &Theory\tablenotemark[2] \\
\tableline
cubic   & $a$ (\AA)       &3.996                   &3.948 \\
tetrag. & $a$ (\AA)       &3.992                   &3.938 \\
        & $c$ (\AA)       &4.036                   &3.993 \\
        & $\delta_{\rm x}$(Ti)             &0.0135  &0.0128 \\
        & $\delta_{\rm x}$(O$_{\rm I}$)     &-0.0150 &-0.0150 \\
        & $\delta_{\rm x}$(O$_{\rm III}$)    &-0.0240 &-0.0232 \\
\end{tabular}
\tablenotetext[1] {Ref.\ \onlinecite{mitsui}.}
\tablenotetext[2] {Ref.\ \onlinecite{king}.}
\label{tab:bulk_par}
\end{table}

BaTiO$_3$ undergoes a series of phase transitions as the temperature
is reduced, from the high-symmetry paraelectric cubic phase to
FE phases with tetragonal, orthorhombic, and rhombohedral unit
cells.  The tetragonal structure is of special interest, since
it is the room-temperature structure.  In this paper, we are thus
primarily interested in the surfaces of the room-temperature
tetragonal phase, although for comparison we also present results
for surface of the elevated-temperature cubic phase.  Ideally one
would like to do this by carrying out {\it ab-initio}
molecular-dynamics simulations at the temperatures at which these
phases are stable, but unfortunately this is not practical.
Instead, we carry out ground-state ($T=0$) calculations, but
subject to the imposition of the appropriate (tetragonal or cubic)
symmetry in order to prevent the system from adopting the true
rhombohedral $T=0$ structure.  This is clearly an approximation,
but we think it is a reasonable one.  The computed ground-state
structural parameters for the cubic and tetragonal bulk phases are
given in Table \ref{tab:bulk_par}, together with experimental
values for comparison.  (Again, the theoretical values are for
$T=0$ structures with the appropriate symmetry imposed.) The
computed lattice constants $a$ and $c$ for the cubic and tetragonal
phases are $1-2\%$ smaller than the experimental ones; this
underestimation is typical of LDA calculations.  We use the
theoretical unit cell parameters in all calculations presented
here.

\begin{figure}
\epsfxsize=2.0in
\centerline{\epsfbox{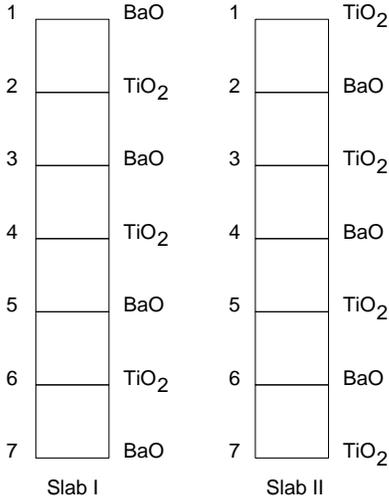}}
\medskip
\caption{
Schematic arrangement of layers in the BaO-terminated (slab I)
and TiO$_2$-terminated (slab II) supercell geometries.  Layers
1 and 7 are surface layers.
\label{fig1}}
\end{figure}

As shown schematically in Fig.~1, the periodic slab corresponding
to the type-I (BaO terminated) surface contains 17 atoms (4 BaO
layers and 3 TiO$_2$ layers).  Similarly, the type-II (TiO$_2$
terminated) slab contains 18 atoms (4 TiO$_2$ layers and 3 BaO
layers).  For both cases, the slabs were thus three lattice
constants thick; the vacuum region was two lattice constants
thick.  The $z$-axis is taken as normal to the surface, and the
$M_z$ mirror symmetry with respect to the central layer
was imposed in all cases.  For surfaces of the cubic phase,
mirror symmetries $M_x$ and $M_y$ were also preserved.
For the tetragonal case, the polarization vector and thus
the tetragonal $c$-axis were chosen to lie along $x$ (parallel
to the surface); here, $M_y$ symmetry was respected but $m_x$
symmetry was allowed to be broken.  As mentioned earlier, the
choice of polarization parallel to the surface is motivated by the
fact that no charge accumulation results at the surface for this
case, and thus no depolarization fields appear.\cite{zhong1}
The case of the tetragonal $c$-axis lying perpendicular to
the surface was considered by Cohen.\cite{cohen1,cohen2} 

The calculations were done using a (4,4,2) Monkhorst-Pack
mesh.\cite{mesh}  This corresponds to three and four k-points in
the irreducible Brillouin zone (BZ) for the cubic and tetragonal
supercells, respectively.  To test the convergence with respect to
k-point sampling, we repeated the calculation for a (6,6,2) mesh,
finding that the surface energy differed by less than $3\%$.  When
the vacuum region was enlarged to three layers in thickness, the
surface  energy changed by less than $4\%$.

In order to study the relative stability of the two kinds of
surface terminations, it is necessary to introduced appropriate
chemical potentials.\cite{guo} To simplify our analysis, we think
of the independent constituents of the slab as being BaO and
TiO$_2$ units.  We define $E_{\rm f}$ to be the {\it formation
energy} needed to make bulk BaTiO$_3$ from BaO and TiO$_2$, per
formula unit.  Thus we have
\begin{equation}
 -E_{\rm f} = E_{\rm BaTiO_3} - E_{\rm BaO} - E_{\rm TiO_2} 
\label{Ef}
\end{equation}
(by convention, $E_{\rm f}>0$).  $E_{\rm BaTiO_{3}}$, $E_{\rm BaO}$,
and $E_{\rm TiO_{2}}$ are the energies of the bulk crystals, per
formula unit, measured relative to isolated ion cores and
electrons. BaTiO$_3$ was calculated in the relaxed tetragonal
structure, and TiO$_2$ in the relaxed rutile structure.

Now, we define the two chemical potentials $\mu_{\rm TiO_{2}}$ and 
$\mu_{\rm BaO}$ in such a way that $\mu_{\rm TiO_{2}}=0$
corresponds to a system in contact with a reservoir of bulk
crystalline TiO$_2$, and similarly for $\mu_{\rm BaO}$.
Furthermore, if we insist that the system is always in equilibrium
with a reservoir of bulk BaTiO$_3$, then we have that
\begin{equation}
 \mu_{\rm BaO} + \mu_{\rm TiO_2} = -E_{\rm f} 
\label{musum}
\end{equation}
Thus, only one of $\mu_{\rm BaO}$ and $\mu_{\rm TiO_{2}}$ is an
independent degree of freedom. We arbitrarily chose $\mu_{\rm
TiO_{2}}$ as the independent one.  Then $\mu_{\rm TiO_{2}}$ can be
allowed to vary over the range
\begin{equation}
 -E_{\rm f} \le \mu_{\rm TiO_{2}}\le 0 \;.
\label{range}
\end{equation}
At $\mu_{\rm TiO_{2}}=-E_{\rm f}$ the system is in
equilibrium with BaO and BaTiO$_3$, and for lower values bulk
crystallites of BaO can precipitate.  Similarly, above
$\mu_{TiO_2}=0$, bulk crystallites of TiO$_2$ can form.

\begin{table}
\caption{Atomic relaxations of the Ba-terminated surface (slab I)
in the cubic (C) and tetragonal (T) phases, given as a fraction of
$a$ or $c$, with respect to ideal positions (i.e., for
$\delta_x$, with respect to the $M_x$ symmetry plane).}
\vspace{0.2in}
\begin{tabular}{lrrr}
Atom   &$\delta_z$ (C) &$\delta_x$ (T) &$\delta_z$ (T) \\
\tableline
Ba(1)     &-0.0279  &-0.0142 &-0.0277 \\
O$_{\rm III}$(1)  &-0.0140  &-0.0298 &-0.0126 \\
Ti(2)     &0.0092   &-0.0086 &0.0098  \\
O$_{\rm I}$(2) &0.0048   &-0.0297 &0.0059 \\
O$_{\rm II}$(2)&0.0048   &-0.0240 &0.0045 \\
Ba(3)     &-0.0053  &-0.0149 &-0.0059 \\
O$_{\rm III}$(3)  &-0.0026  &-0.0280 &-0.0020 \\
Ti(4)     &0      &-0.0034 &0     \\
O$_{\rm I}$(4) &0      &-0.0340 &0 \\
O$_{\rm II}$(4)&0      &-0.0256 &0 \\
\end{tabular}
\label{tab:slabI}
\end{table}

Therefore, the grand thermodynamic potential per surface unit
cell is given by
\begin{eqnarray}
 F = {1\over 2} [ E_{\rm slab}
       - N_{\rm TiO_2} && (\mu_{\rm TiO_2} + E_{\rm TiO_2}) \nonumber \\
       && - N_{\rm BaO}(\mu_{\rm BaO} + E_{\rm BaO}) ]
\label{F}
\end{eqnarray}
(the factor of 1/2 appearing because the cell contains
two surfaces), where $E_{\rm slab}$ is the energy of the relaxed
slab in the tetragonal phase.  For example, for the type-I
slab, one has $N_{\rm TiO_{2}}=3$ and $N_{\rm BaO}=4$.
Eqs.\ (\ref{musum}) and (\ref{F}) give $F$ as a function
of $\mu_{\rm TiO_{2}}$ over the range of Eq.\ (\ref{range}).

\section{Results and Discussions}
\label{sec:results}      

\subsection{Structural relaxations}
\label{sec:structure}      

First we determined the equilibrium atomic positions for our two
types of slabs in the two phases.  For the cubic surface, we
started from the ideal structure and relaxed.  For the tetragonal
surface, we obtained a starting guess by superposing the
$z$-displacements from the relaxed cubic surface with
$x$-displacements from the bulk tetragonal structure.\cite{king}
The relaxed geometries are summarized in Tables \ref{tab:slabI} and
\ref{tab:slabII}.  In these tables, the atoms are listed in the
same order as shown in Fig.~1.  (Coordinates are only listed for
atoms in the top half of the slab, $z\ge0$; the others are
determined by the $M_z$ mirror symmetry.)  By symmetry, there are
no forces along $\hat x$ or $\hat y$ for the cubic surfaces, and no
forces along $\hat y$ for the tetragonal surface.
Also due to the crystal termination, the two O atoms associated
with the Ti atom (O$_{\rm I}$ and O$_{\rm II}$), are no longer
equivalent in the tetragonal phase (I, II, and III indicate
the O that is connected to Ti by a bond along $x$, $y$, and $z$,
respectively).

\begin{table}
\caption{Atomic relaxations of the Ti-terminated surface (slab II)
in the cubic (C) and tetragonal (T) phases, given as a fraction of
$a$ or $c$, with respect to ideal positions.}
\vspace{0.2in}
\begin{tabular}{lrrr}
Atom   &$\delta_z$ (C) &$\delta_x$ (T) &$\delta_z$ (T) \\
\tableline
Ti(1)     &-0.0389 &0.0005  &-0.0331 \\
O$_{\rm I}$(1)  &-0.0163 &-0.0499 &-0.0100 \\
O$_{\rm II}$(1) &-0.0163 &-0.0366 &-0.0071 \\
Ba(2) &0.0131  &-0.0148 &0.0186 \\
O$_{\rm III}$(2) &-0.0062 &-0.0292 &-0.0023 \\
Ti(3)        &-0.0075 &0.0019  &-0.0058 \\
O$_{\rm I}$(3) &-0.0035 &-0.0372 &-0.0022 \\
O$_{\rm II}$(3)  &-0.0035 &-0.0278 &-0.0023 \\
Ba(4)          &0     &-0.0111 &0 \\
O$_{\rm III}$(4) &0     &-0.0276 &0 \\
\end{tabular}
\label{tab:slabII}
\end{table}

From Tables \ref{tab:slabI} and \ref{tab:slabII}, we can see that
the largest relaxations are on the surface-layer atoms, as
expected.  They are especially important for the Ti atoms in the
Ti-terminated slabs, in the direction perpendicular to the surface.
This can plausibly be explained by noting that for bulk BaTiO$_3$,
the filled Ba levels lie well below the oxygen-2$p$ valence-bands
and do not hybridize strongly, so that the Ba atom is a relative
spectator in the bonding.\cite{king}
In the tetragonal case, we can also see that the asymmetry
between the two O atoms
lying in a common surface plane is significant.  Bond lengths on the
surface change by less than 1.5\% with respect to the bulk in the
same phase (the latter values are the experimental ones, taken from
Mitsui {\it et al.}\cite{mitsui}).

We computed the average displacements $\beta$ and the
rumpling $\eta$ for the surface layers; the results are
given in Table \ref{tab:relax}.
To fix notation, let $\delta z(\rm M)$ be the change in the
surface-layer metal-atom $z$ position relative to the ideal
unrelaxed structure, and $\delta z(\rm O)$ be the same for the
surface oxygens (defined as $[\delta z(\rm O_I) + \delta z(\rm
O_{II})]/2$ for a TiO$_2$ layer).  Then the
surface relaxation parameter $\beta$ is defined as 
$[\delta z(\rm M) + \delta z(\rm O)]/2$, and the rumpling $\eta$
is defined as $[\delta z(\rm O)-\delta z(\rm M)]/2$.  
We find that the surface layers contract substantially inwards
towards the bulk, with both the metal and oxygen ions relaxing
in the same direction.

\begin{table}
\caption{Calculated interlayer relaxation ($\beta$) and rumpling
($\eta$) for the surface layer of the relaxed slabs in the cubic (C)
and tetragonal (T) phases (\AA) .}
\vspace{0.2in}
\begin{tabular}{lrrrr}
Slab       &$\beta$ (C)  &$\eta$ (C) &$\beta$ (T) &$\eta$ (T) \\
\tableline
Slab I     &-0.08 &0.03 &-0.08 &0.03 \\ 
Slab II    &-0.11 &0.05 &-0.08 &0.05 \\
\end{tabular}
\label{tab:relax}
\end{table}

Cohen\cite{cohen2} has computed the surface relaxations of
an asymmetrically-terminated cubic slab (BaO on one surface and
TiO$_2$ on the other).  Detailed quantitative agreement
is probably not expected, because (i) only the surface-layer
atoms were relaxed in Cohen's calculation, and (ii) the
asymmetric termination introduces a small electric field which
may have influenced the relaxations.  Nevertheless, we do find
qualitative agreement.  Cohen finds that the Ba and O atoms
relax inwards by 0.043 and 0.033 lattice constants, respectively,
on the Type-I surface; and the Ti and O atoms relax inwards by
0.048 and 0.027, respectively, on the type-II surface.
These can be compared with the first two entries in the
$\delta_z$(C) column of Tables \ref{tab:slabI} and \ref{tab:slabII}.
It can be seen that Cohen's values are 20-100\% larger in magnitude,
but of the same sign, as those that we calculate.  Similarly,
Cohen computes values of -0.15\AA\ for the average surface layer
relaxation, to be compared with the values given by us in the
column $\beta$(C) of Table \ref{tab:relax}.  The rumpling
computed by Cohen is also of the same sign, but different in
detail, as that calculated by us.

We are not aware of any experimental surface structure
determination for BaTiO$_3$ with which we can compare our theory.
However, we note that a LEED I-V study of the corresponding
SrTiO$_3$ surfaces\cite{hikita} comes to an opposite conclusion,
suggesting an outward relaxation of the surface layer on the order
of 0.1\AA.  While we have not carried out a parallel calculation on
the SrTiO$_3$ surface, it appears unlikely that the mere replacement
of Sr by the chemically similar Ba could be responsible for such a
large qualitative discrepancy.  Thus, we suggest that the
experimental interpretation\cite{hikita} should be reexamined.

\subsection{Influence of the surface upon ferroelectricity}
\label{sec:FE}

It is important to understand whether the presence of the surface has
a strong effect upon the near-surface ferroelectricity.  For example,
is the FE order enhanced near the surface, or is it suppressed?
As we shall see in Sec.\ \ref{sec:surfen}, the energy scale of
the surface relaxations is larger than the energy scale of the
FE double-well potential.  Thus, a strong effect is possible.
To analyze whether it really occurs, we computed an average FE
distortion $\delta_{\rm FE}$ for each layer of the slab.  We define
$\delta_{\rm FE} =\delta_x({\rm Ba})-\delta_x(\rm O_{III})$
for a BaO plane, and
$\delta_{\rm FE} =\delta_x({\rm Ti})
                -[\delta_x{\rm(O_I)}+\delta_x{\rm(O_{II})}]/2$
for a TiO$_2$ plane.

\begin{table}
\caption{Calculated FE distortion $\delta_{\rm FE}$ of the relaxed
slabs, for each layer (units of lattice constant).  Last line gives
theoretical bulk values for reference.}
\vspace{0.2in}
\begin{tabular}{lllll}
&
\multicolumn{2}{c}{Slab I} &
\multicolumn{2}{c}{Slab II} \\
Layer & $\delta_{\rm FE}$(BaO) & $\delta_{\rm FE}$(TiO$_2$)
      & $\delta_{\rm FE}$(BaO) & $\delta_{\rm FE}$(TiO$_2$) \\
\tableline  
1    & 0.0131 &        &        & 0.0345 \\
2    &        & 0.0264 & 0.0165 &        \\
3    & 0.0157 &        &        & 0.0434 \\
4    &        & 0.0259 & 0.0144 &        \\
Bulk & 0.0232 & 0.0278 & 0.0232 & 0.0278 \\
\end{tabular}
\label{tab:fe_dist}
\end{table}

Our calculated values for $\delta_{\rm FE}$ are given in Table
\ref{tab:fe_dist}; the last row of the table gives the bulk values
for reference.  No clear pattern appears to emerge from these
results, although we do note a moderate enhancement of the FE
instability in the TiO$_2$ layers for the TiO$_2$-terminated
surface.
 
The lack of a clear trend for the influence of surface effects
upon the FE distortion can be understood, at least in part, by
noting that the FE mode is only one of three zone-center modes
having the same symmetry.\cite{king}  The FE mode is distinguished
as the one that is soft ($\omega^2<0$) in the cubic structure,
but there is no particular reason why the surface relaxation should
couple more strongly to this mode than to the others.  We have
estimated how strongly the surface relaxations are related to
each of the zone-center modes by the following procedure.
We calculate the forces for a tetragonal surface slab in which the
displacements in the $x$ direction are those of the ideal bulk
tetragonal structure, while the displacements in the $z$ direction
are taken from the relaxed cubic surface.
The forces in the $x$ direction are then projected onto each of the
zone-center bulk modes polarized along $x$.  That is, the force
for each type of atom (e.g., O$_{\rm III}$) was summed over all such
atoms in the slab, and the inner product was then taken between the
resulting force vector and the bulk mode eigenvectors.

We found that the FE mode accounts for only about 31\% and 26\% of this
force vector for the type-I and type-II slabs respectively.  Thus,
it seems that the distortions induced by the presence of the surface
are mostly of non-FE character, helping to explain why the FE order
is not as strongly affected as might have been guessed.

\subsection{Surface energies}
\label{sec:surfen}

We turn now to a study of the surface energetics.  Following the
approach outlined in Sec.~\ref{sec:theory}, we calculated the
grand thermodynamic potential $F$ for our two types of surface
as a function of TiO$_2$ chemical potential.  The results are shown in
Fig.~2.   In order to attain optimal cancelation of errors,
$E_{\rm TiO_2}$ and $E_{\rm BaO}$ were calculated within the LDA
using the same pseudopotentials, and with the same 25 Ry energy cut-off.
A similar k-point sampling as for the surfaces of BaTiO$_3$
was also employed.  In this way, we obtained $E_{\rm f}=3.23$ eV
for the formation energy of BaTiO$_3$.
This quantity fixes the range of physical values of $\mu_{\rm TiO_2}$;
the left and right boundaries of Fig.~2 correspond to a system in
thermodynamic contact with bulk BaO and bulk TiO$_2$ respectively.
It can be seen that both surfaces have a comparable range of
thermodynamic stability, indicating that either type-I or type-II
surfaces could be formed depending on whether growth occurs
in Ba-rich or Ti-rich conditions.

\begin{figure}
\epsfxsize=2.6in
\centerline{\epsfbox{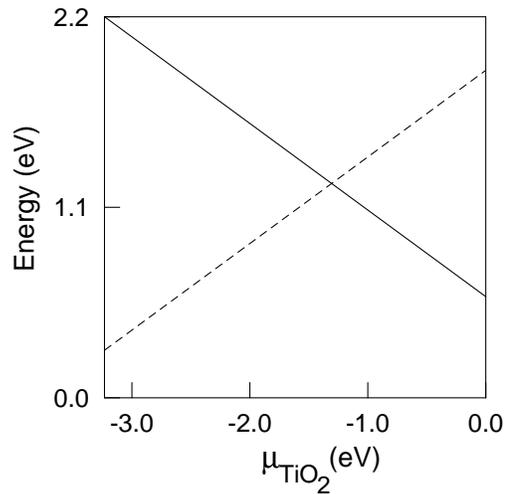}}
\medskip
\caption{
Grand thermodynamic potential $F$ as a function of the chemical
potential $\mu_{\rm TiO_{\rm 2}}$,
for the two types of surfaces, in the tetragonal phase.
Dashed and solid lines correspond to type-I (BaO-terminated)
and type-II (TiO$_2$-terminated) surfaces respectively.
\label{fig2}}
\end{figure}

The {\it average} surface energy $E_{\rm surf}$ (i.e., the average
of $F$ for the two kind of surfaces) is independent of
$\mu_{\rm TiO_2}$.  Therefore, this quantity is suitable for
comparisons.  For the cubic phase, we estimated $E_{\rm surf}$
for the (001) surfaces to be $1.241$ eV per surface unit cell
($1265$ erg/cm$^2$); and for the tetragonal phase, it was estimated
to be $1.237$ eV per surface unit cell ($1260$ erg/cm$^2$).  The
value of the average $E_{\rm surf}$ calculated in
Ref.~\onlinecite{cohen2}
for the symmetrically-terminated cubic (001) surfaces is
$920$ erg/cm$^2$.  As pointed out in that paper, the large value
of $E_{\rm surf}$ may help explain why BaTiO$_3$ does not cleave
easily, but fractures instead.

In order to compute the surface relaxation energy $E_{\rm relax}$,
we computed the average surface energy $E_{\rm unrel}$ for the
{\it unrelaxed} cubic slabs (i.e., atoms in the ideal cubic
perovskite positions), using the same k-point sampling as for
the relaxed systems.  We obtained $E_{\rm unrel}=1.358$ eV.
Thus, the relaxations account for $0.127$ eV of the surface energy
per surface unit cell (or about $130$ erg/cm$^2$).

Note that $E_{\rm relax}$ is many times larger than the bulk
ferroelectric well depth, estimated to be of the order of $0.03$ eV.
This would indicate that the surface is capable of acting as a
strong perturbation on the FE order.  As explained in
Sec.~\ref{sec:FE}, however, the actual effect is more modest than
one would guess based on energetic considerations alone.

\subsection{Surface band structure}

We have carried out LDA calculations of the surface electronic
structure for our various surface slabs.  While the LDA is
well known to be quantitatively unreliable as regards excitation
properties such as band gaps, we believe that the results presented
here are nevertheless likely to be qualitatively correct.
The bulk band gap in our calculation is 1.8 eV, to be compared with
the experimental value of 3.2 eV; this level of disagreement is
typical for the LDA.

\begin{figure}
\epsfxsize=3.2in
\centerline{\epsfbox{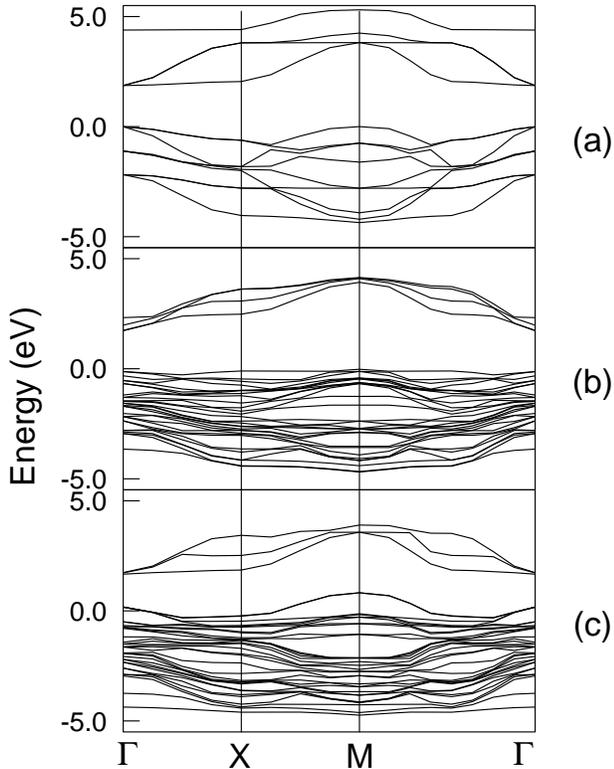}}
\medskip
\caption{
Calculated band structures for BaTiO$_3$ in the cubic phase.
(a) Surface-projected bulk band structure.
(b) BaO-terminated surface (slab I).
(c) TiO$_2$-terminated surface (slab II).
The zero of energy corresponds to the bulk valence band maximum.
Only the lowest few conduction bands are shown.
\label{fig3}}
\end{figure}

Fig.~3 shows the calculated LDA band structure for cubic bulk
BaTiO$_3$ projected onto the surface BZ, and the surface band
structures for the Ba- and Ti-terminated relaxed surfaces in the
cubic phase.
(The zero of energy for each surface slab was established by
aligning the Ba or Ti semicore $s$ states in the interior layers
of the slab with those of the bulk.)
Plots for the tetragonal surfaces would look similar, except that
the tendency for states to intrude into the gap is stronger for
the cubic case.  This can be seen in Table \ref{tab:gaps}, where
we list the calculated band gaps for both cubic and tetragonal
slabs.  We therefore tend to focus on the cubic surfaces, where
it is easier to identify and characterize the surface states.

\begin{table}
\caption{Calculated band gaps for relaxed cubic (C) and
tetragonal (T) surface slabs (eV).}
\vspace{0.2in}
\begin{tabular}{lcc}
Slab             & C      & T    \\
\tableline  
Slab I           & 1.80   & 2.01 \\
Slab II          & 0.84   & 1.18 \\
Bulk             & 1.79   & 1.80 \\
\end{tabular}
\label{tab:gaps}
\end{table}

First, we can see that on the Ba-terminated (type-I) surface,
the gap is not reduced and there are no deep gap states.  On the
Ti-terminated (type-II) surface, however, the gap is reduced,
especially for the cubic case.  As can be seen from Fig.~3(c),
there is a tendency for valence-band states to intrude upwards
into the lower part of the band gap for this surface, especially
near the $M$ point of the surface BZ.  (Qualitatively similar
results can be seen in Fig.\ 3 of Ref.\ \onlinecite{cohen2}.)
However, the conduction band does not change much with respect
to the bulk.  Moreover, we do not see signs of any true ``deep-gap''
states lying near the center of the gap.  As noted earlier, the
existence of such deep-gap states remains controversial
experimentally.  Our work suggests that if gap states do exist
in connection with non-defective (001) surfaces, it is likely
that they would be found in the lower part of the band gap, and
that this would be indicative of exposed TiO$_2$ (as opposed to
BaO) surface planes.

\begin{figure}
\epsfxsize=2.2in
\centerline{\epsfbox{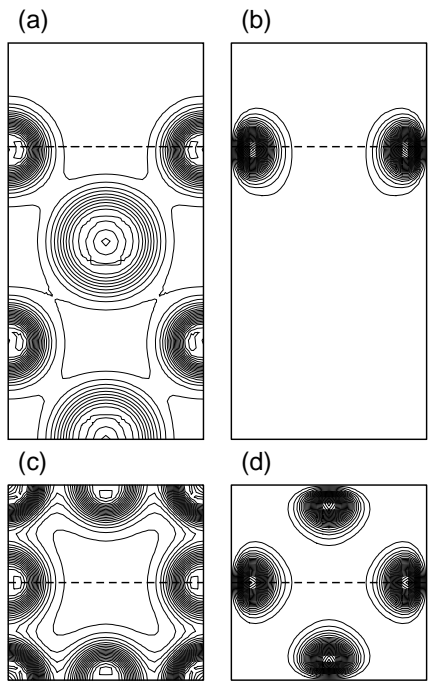}}
\medskip
\caption{
Charge-density plots for relaxed Ti-terminated surface (slab II) in the
cubic phase.  Panels (a-b) show cuts on a vertical $x$-$z$ BaO plane;
panels (c-d) show cuts on the $x$-$y$ surface TiO$_2$ plane.
Total charge density, (a) and (c); charge density of highest
occupied state at $M$ point of the surface BZ, (b) and (d).
\label{fig4}}
\end{figure}

Figure 4 illustrates the character of the valence-band state at
the $M$ point that is intruding into the lower part of the gap.
The total charge density is also shown for reference.  It
can be seen that this state is composed of O $2p$ lone-pair
orbitals lying in the surface plane.  Further inspection
shows that the special feature of this state is that the
wavefunction has four nodal planes [(100), (110), (010), and
($\bar 1$10)] intersecting at the Ti sites.  This precludes the
presence of any Ti $3d$ character (in fact, any Ti character of
angular momentum $l<4$).  In the bulk, the oxygen $2p$ orbitals
are all hybridized with Ti $3d$ orbitals to some extent, and the
level repulsion associated with this hybridization pushes the
energy location of the valence-band states downward in energy.
Thus, the energy of the unhybridized O $2p$ lone-pair 
surface state at the $M$ point is left intruding into the lower
part of the gap.

This insight makes it possible to understand other features of
the surface band structures as well.  For example, on the
BaO-terminated surface, every surface oxygen atom is directly
above a Ti atom, and is strongly hybridized to it.  Thus, there
is no such tendency for the formation of surface states in this
case.  Returning to the TiO$_2$-terminated surface, there seems
to be a weaker tendency for the intrusion of a valence-band
derived surface state at $\Gamma$.  This state turns out to have
a single nodal plane passing through the Ti site, so that
hybridization with Ti $3d$ orbitals is only weakly allowed (by
the breaking of $M_z$ mirror symmetry across the surface plane).

We expect that these results would remain qualitatively valid for
other II-VI perovskites such as SrTiO$_3$ or PbZrO$_3$.
For I-V perovskites such as KNbO$_3$ and LiTaO$_3$, however,
the (001) surface is non-stoichiometric, and the surface electronic
structure would be expected to be quite different.

\section{Summary} 

In summary, we have carried out LDA density-functional calculations of
BaO- and TiO$_2$-terminated (001) surfaces for cubic and tetragonal
phases of BaTiO$_3$.  By minimizing the forces on the ions,
we obtained the relaxed ionic structures.  As would have been expected
from the bulk electronic levels of BaTiO$_3$,  the most important
relaxations occur for the TiO$_2$-terminated surfaces.  There
appears to be a modest tendency for the surface relaxations to enhance
the FE distortion on that surface, although the situation is
complicated by the fact that the relaxations excite modes other than
the soft zone-center one.  

The free energies for the different surfaces were calculated as
a function of the TiO$_2$ chemical potential.  In particular,
the average surface energy was found to be about $1260$ erg/cm$^2$.
The surface relaxation energies were found to be around 10\% of the
total surface energy.

In accord with previous theoretical reports, no mid-gap surface levels
are found.  But for the TiO$_2$ terminated surfaces, there is a
substantial reduction of the bulk gap, especially for the cubic-phase
surface.  This reduction results from the intrusion of states of
valence-band character into the lower part of the band gap, especially
near the M point of the surface BZ.

\acknowledgments

This work was supported by the ONR grant N00014-91-J-1184 and NSF
grant DMR-91-15342.
We also acknowledge Cray C90 computer time provided by the
Pittsburgh Supercomputing Center under grant DMR930042P.

\end{document}